\documentclass[preprint,prb,showpacs]{revtex4}
\usepackage{amssymb,graphicx}

\begin{document}
\title{Elasticity and piezoelectricity of zinc oxide crystals, single layers, and possible single-walled nanotubes}

\author{Z. C. Tu}
\affiliation{Computational Materials Science Center, National
Institute for Materials Science, Tsukuba 305-0047, Japan}
\author{X. Hu}
\affiliation{Computational Materials Science Center, National
Institute for Materials Science, Tsukuba 305-0047, Japan}
\begin{abstract}
The elasticity and piezoelectricity of zinc oxide (ZnO) crystals
and single layers are investigated from the first-principles
calculations. It is found that a ZnO thin film less than three
Zn-O layers prefers a planar graphite-like structure to the
wurtzite structure. ZnO single layers are much more flexible than
graphite single layers in the elasticity and stronger than boron
nitride single layers in the piezoelectricity. Single-walled ZnO
nanotubes (SWZONTs) can exist in principle because of their
negative binding energy. The piezoelectricity of SWZONTs depends
on their chirality. For most ZnO nanotubes except the zigzag type,
twists around the tube axis will induce axial polarizations. A
possible scheme is proposed to achieve the SWZONTs from the
solid-vapor phase process with carbon nanotubes as
templates.
\end{abstract}
\maketitle

\section{Introduction}
Zinc oxide (ZnO) materials have attracted extensive attention for
half a century because of their excellent performance in optics,
electronics and photoelectronics. \cite{ozgur} They are important
for low cost productions of green, blue-ultraviolet, and white
light-emitting devices due to their wide band gap ($\sim$3.37 eV)
and large exciton binding energy ($\sim$60 meV). They can also be
used as sensors and transducers owing to their strong
piezoelectricity. Recently, many ZnO nanostructures have been
synthesized in experiments. \cite{wangzlmt} Among them, the
quasi-one-dimensional structures have become the leading research
objects because of their novel chemical, electrical, mechanical
and optical properties. \cite{wangzljpc} The ultraviolet lasing in
ZnO nanowires \cite{Huangmhsci,Hauptmjap} and strong
photoluminescence in ZnO nanorods \cite{Parkwiapl} have been
demonstrated. Nanobelts, nanorings and nanohelixes are also
synthesized, \cite{Panzw,Kongxynl,Hugheswl,Kongxysci} which may be
useful for field-effect transistors. \cite{Arnoldmsjpcb}

Since the discovery of carbon nanotubes in 1991, \cite{Iijima} the
synthesis of tubular nanostructures has raised worldwide interest.
A lot of inorganic nanotubes, such as ZrS$_2$, NbSe$_2$, SiO$_2$,
TiO$_2$, BN nanotubes, etc. are achieved by several groups.
\cite{Rao2003} Researchers have also tried to fabricate ZnO
nanotubes through various methods including thermal reduction,
\cite{Hujqcm} vapor phase growth,
\cite{Xingyjapl,Xingyjssc,Wangrmnjp} hydrothermal growth,
\cite{Sunyam} vapor-solid process, \cite{Kongxyjpcb} sol-gel
template process, \cite{Gugsssc} plasma-assisted molecular beam
epitaxy, \cite{Yanjfjcg,Zhangxhjpcb} pyrolysis of zinc
acetylacetonate, \cite{Wujjapl} and Zn(NH$_3$)$_4^{2+}$ precursor
thermal decomposition. \cite{Zhangjcc} All ZnO nanotubes thus
obtained have the wurtzite structure with diameter of 30--450 nm
and thickness of 4--100 nm. Much effort has been made to realize
much thinner and smaller ZnO nanotubes. A natural question is: Can
we manufacture single-walled ZnO nanotubes (SWZONTs)? In this
paper, We will answer this question based on the first-principles
calculations (the ABINIT package \cite{Gonzexcms}) and the
experimental methods
\cite{Xingyjapl,Xingyjssc,Wangrmnjp,Kongxyjpcb} for synthesizing
ZnO nanobelts and nanotubes with the wurtzite structure.
Additionally, we will investigate the elasticity and
piezoelectricity of ZnO crystals and single layers within the
framework of density-functional theory (DFT) \cite{Kohnw} and
density-functional perturbation theory (DFPT).
\cite{Gonzexpr,Baronirmp,Hamannprb} The piezoelectricity of
single-walled ZnO nanotubes is also derived from the
piezoelectricity of ZnO single layers.

The rest of this paper is organized as follows: In
Sec.~\ref{crystal}, we present the results of elastic constants
and piezoelectric coefficients of ZnO crystals. We also compare
them with the experimental results and previous theoretical
results in literatures. The main aim of this section is to test
whether our computational procedure can validly describe the
physical properties of the ZnO system or not. In Sec.~\ref{layer},
we optimize a Zn-O single layer cut from the wurtzite crystal and
arrive at a planar graphite-like structure. Comparing the binding
energy between the crystal and the planar single layer, we deduce
that a ZnO thin film less than three Zn-O layers prefers a planar
graphite-like structure to the wurtzite structure. The elastic and
piezoelectric constants of ZnO single layers are also calculated.
In Sec.~\ref{nanotube}, we calculate and optimize different
SWZONTs. The SWZONTs are shown to exist in principle because of
their negative binding energy. Their piezoelectricity depends on
the chirality, and for most of them except the zigzag type, twists
around the tube axis will induce axial polarizations. We also
propose a possible scheme to realize the SWZONTs from the
solid-vapor phase process with carbon nanotubes as templates. In
Sec.~\ref{conclusion}, we give a summary and propose some
potential applications of SWZONTs.

\section{Zinc oxide crystals\label{crystal}}
The ground state of a ZnO crystal has a wurtzite structure (space
group P6$_3$mc) with two zinc and two oxygen atoms per unit cell.
The lattice constants and internal parameter obtained from
experiments are as follows: $a=b=3.250$\AA, $c= 5.204$\AA, and
$u=0.382$, respectively.\cite{Karzelprb}

We optimize the structure parameters and calculate the elastic and
piezoelectric constants of ZnO crystals. The calculations are
carried by taking Troullier-Martins pseudopotentials,
\cite{Troullier} plane-wave energy cutoff (ecut) of 60 Hartree,
and $6\times6\times3$ Monkhorst-Pack k-points \cite{MonkhorstPack}
in Brillouin-zone. The exchange-correlation energy are treated
within the local-density approximation in the Ceperley-Alder form
\cite{CeperleyAlder} with the Perdew-Wang parametrization.
\cite{PerdewWang}

Using the experimental parameters to construct an initial unit
cell, we have optimized the structure parameters and summarized
them in Table~\ref{strutpara}. Results by other groups are also
listed for comparison. It is easy to see that our DFT results are
quite close to those of Wu \emph{et al.} and agree well with the
experimental results (errors in 2\%).

Adopting the optimized structure, we can calculate the elastic
constants and piezoelectric coefficients of ZnO crystals. The
elastic constants reflect the stress-strain relation of materials.
In terms of the symmetry of ZnO crystals (wurtzite structure),
this relation can be expressed in the matrix form: \cite{Nyebook}
\begin{equation}
\left[\begin{array}{c} \sigma_1\\ \sigma_2\\
\sigma_3\\
\sigma_4\\
\sigma_5\\
\sigma_6\end{array}\right]=\left[\begin{array}{cccccc} c_{11} & c_{12} &c_{13}&0 &0 &0\\
c_{12}&c_{11}&c_{13}&0&0&0\\
c_{13}&c_{13}&c_{33}&0&0&0\\
0&0&0&c_{44}&0&0\\
0&0&0&0&c_{44}&0\\
0&0&0&0&0&\frac{c_{11}-c_{12}}{2}\end{array}\right]\left[\begin{array}{c} \epsilon_1\\ \epsilon_2\\
\epsilon_3\\ \epsilon_4\\ \epsilon_5\\
\epsilon_6\end{array}\right],
\end{equation}
where $\sigma_i$ and $\epsilon_i$ (i=1,$\cdots$,6) represent the
stresses and strains, respectively. There are only five
independent elastic constants: $c_{11}$, $c_{12}$, $c_{13}$,
$c_{33}$, and $c_{44}$. Their values in the present work are
listed in Table~\ref{elaconst}.

Similarly, the piezoelectricity can be expressed in the matrix
form: \cite{Nyebook}
\begin{equation}
\left[\begin{array}{c} P_1\\ P_2\\
P_3-P_3^0\end{array}\right]=\left[\begin{array}{cccccc} 0 & 0 & 0 & 0 &e_{15} &0\\
0&0&0&e_{15}&0&0\\
e_{31}&e_{31}&c_{33}&0&0&0\\
\end{array}\right]\left[\begin{array}{c} \epsilon_1\\ \epsilon_2\\
\epsilon_3\\ \epsilon_4\\ \epsilon_5\\
\epsilon_6\end{array}\right],
\end{equation}
where $P_1$, $P_2$, and $P_3$ are three polarization components
along $\textbf{a}$ direction, the direction perpendicular to
$\textbf{a}$ and $\textbf{c}$, and $\textbf{c}$ direction,
respectively. $P_3^0$ is the spontaneous polarization along
$\textbf{c}$ direction. There are only 3 independent piezoelectric
coefficients: $e_{31}$, $e_{33}$, and $e_{15}$. It is necessary to
point out that here the definition of piezoelectricity is
different from that in Ref.~\onlinecite{Nyebook} where the
piezoelectricity reflects the relation between polarizations and
stresses. The piezoelectric coefficients obtained by the present
work are listed in Table~\ref{piezoconst}.

Good agreements have been achieved between the present results and
those in literatures (see Tables ~\ref{elaconst} and
\ref{piezoconst}). Confirming that our computational procedure can
give reasonable results for bulk ZnO systems, we proceed to
analyze ZnO single layers and nanotubes.

\section{Zinc oxide single layers\label{layer}}

\begin{figure}[!htp]
\includegraphics[width=3.5cm]{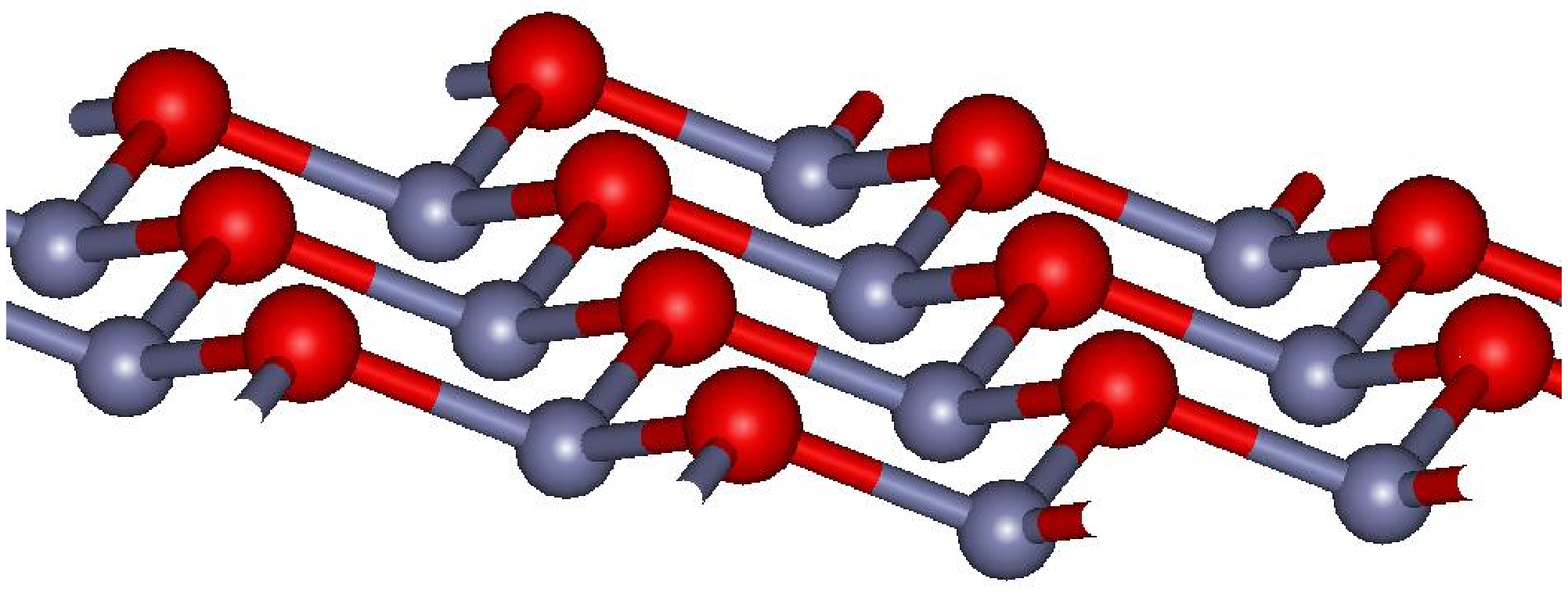}\hspace{1cm}
\includegraphics[width=3.5cm]{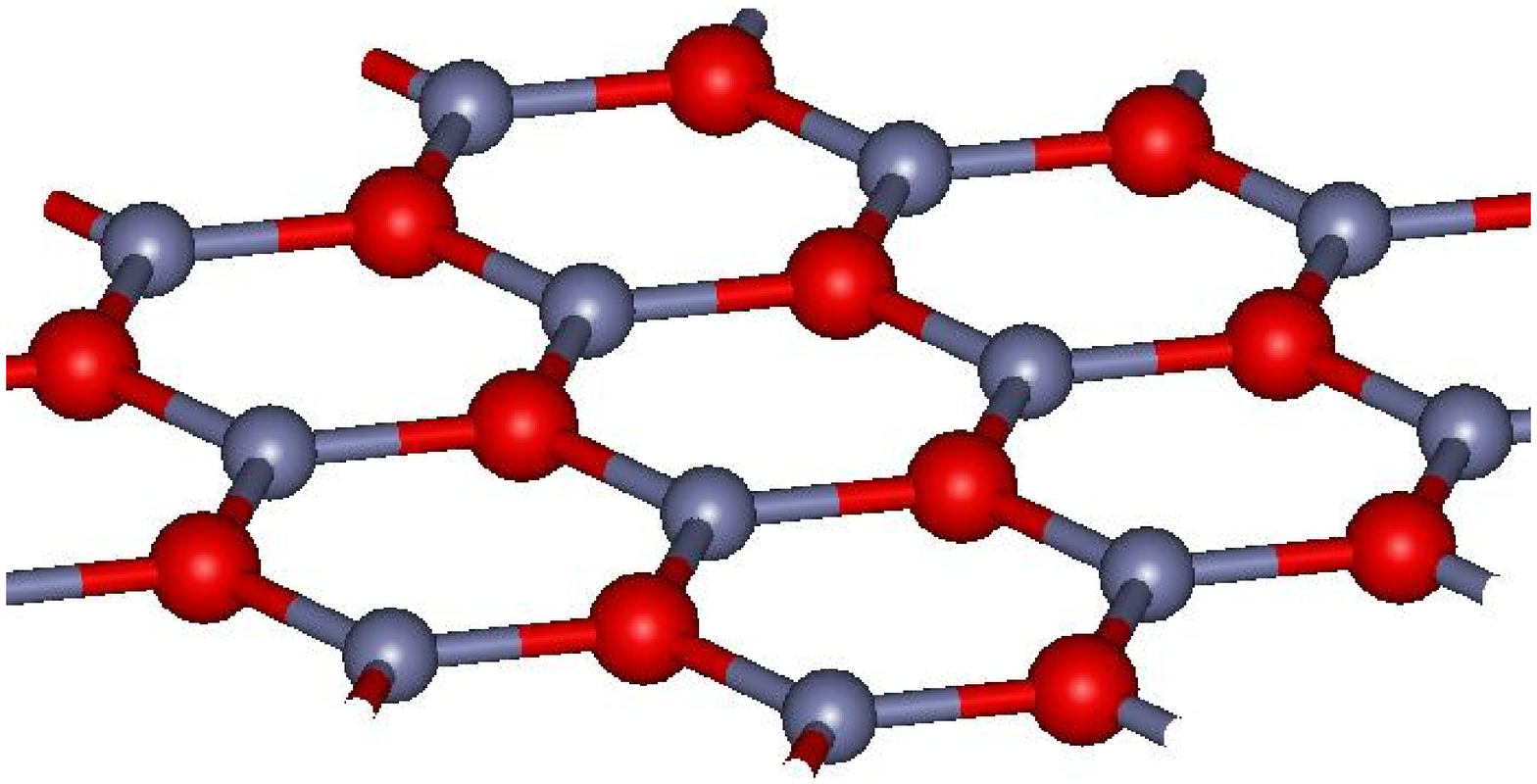}
\caption{\label{singlelay}(Color online) ZnO single layers. The
left side is a single layer taken from the wurtzite ZnO crystal
while the right one is the optimized structure which is a planar
hexagonal lattice. The small and large balls represent zinc and
oxygen atoms, respectively.}\end{figure}

In this section, we discuss the elasticity and piezoelectricity of
ZnO single layers. First, as an initial configuration, we take a
single Zn-O layer (shown in the left side of Fig.~\ref{singlelay})
cut from the wurtzite ZnO crystal, and then optimize its
structure. As an stable configuration, we obtain a planar
graphite-like structure as shown in the right side of
Fig.~\ref{singlelay}. Here we use $6\times6\times1$ Monkhorst-Pack
k-points and two different groups of other input parameters for
calculations. The final optimized results, listed in
Table~\ref{planartab}, are insensitive to the input parameters.
The bond length of the planar structure is slightly smaller than
that of ZnO crystals. Its binding energy is -8.246 eV/ZnO, higher
than the value -8.947 eV/ZnO of ZnO crystals.

Now we consider a wurtzite ZnO thin film with two polar surfaces
($\pm 0001$ faces). Its energy is higher than that of bulk ZnO
with the same number of atoms by $E_c=4.0$ J/m$^2$ (the cleavage
energy density \cite{Wander01}) because of the spontaneous
polarization in \textbf{c} direction. Here we derive the the
critical layer number based on our above result for planar ZnO
single layers. Considering the influence of spontaneous
polarization, we estimate the total binding energy
$E_f=M(-8.947N+E_c\Omega_0)=M(-8.947N+2.21)$ eV for the wurtzite
thin film with $N$ Zn-O single layers, where $\Omega_0$ is the
area for a unit cell of the ZnO crystal in ab plane and $M$ the
number of unit cells in each layer. If we consider the thin film
with $N$ planar single layers and neglect the very weak interlayer
attractions, the total binding energy is about $-8.246MN$ eV. We
find $E_{f}>-8.246MN$ for $N<4$, which implies that a ZnO thin
film less than $N_c=4$ Zn-O layers prefers the planar
graphite-like structure to the wurtzite structure. Here the
estimated value 4 is the lower bound of critical number $N_c$
because we neglect the weak interlayer attractions. Note that the
cleavage energy density may depend on the layer number and
structural relaxation. Considering the structural relaxation,
Meyer and Marx \cite{Meyer2003} found that the the cleavage energy
density varied from 3.0 J/m$^2$ to 3.4 J/m$^2$, weakly dependent
on the the layer number. Using these values, the lower bound of
critical number $N_c$ is changed to 3. If we consider the
interlayer attractions coming from the dipole-dipole interaction
between planar ZnO layers because of their spontaneous
polarizations (the ground state: spontaneous polarizations
antiparallel in the neighbor layers), we should obtain $N_c>3$,
which is the exact meaning of ``the lower bound of critical number
being 3''. Claeyssens \emph{et al.} also obtained a critical
thickness by DFT calculations. \cite{Claeyssens} They investigated
the binding energy of the thin films with $N=4,6,\cdots,24$ single
planar layers (layer distance 2.4 \AA) and compared it with the
binding energy of the thin films (wurtzite structure) with the
same number of Zn-O layers. They found that the ZnO thin film less
than eighteen Zn-O layers prefers the planar graphite-like
structure to the wurtzite structure. As is well known, the widely
used DFT packages cannot accurately describe the interlayer
interactions. \cite{Andersson96,Kohn98} Thus the critical layer
number $N_c=18$ obtained directly from DFT calculations might be
inaccurate although it satisfies $N_c>3$. With the development of
DFT method, one may obtain the accurate critical layer number in
the future when DFT packages can deal well with the interlayer
interactions.

Adopting the optimized structure, we can calculate the elastic and
piezoelectric constants of ZnO single layers. From
Fig.~\ref{graphicly}, we find that the ZnO single layer has the
3-fold rotation symmetry and a reflection symmetry respect to
$x_1$-axis. Thus the stress-strain relation is expressed in the
matrix form: \cite{Nyebook}
\begin{equation}
\left[\begin{array}{c} \sigma_1\\ \sigma_2\\
\sigma_6\end{array}\right]=\left[\begin{array}{ccc} c_{11} & c_{12} &0\\
c_{12}&c_{11}&0\\
0&0&\frac{c_{11}-c_{12}}{2}\end{array}\right]\left[\begin{array}{c} \epsilon_1\\ \epsilon_2\\
\epsilon_6\end{array}\right].
\end{equation}
There are only two independent elastic constants: $c_{11}$ and
$c_{12}$. The piezoelectricity can be expressed as
\begin{equation}
\left[\begin{array}{c} P_1-P_1^0\\ P_2\end{array}\right]=\left[\begin{array}{ccc} e_{11} & -e_{11} &0\\
0&0&-e_{11}\end{array}\right]\left[\begin{array}{c} \epsilon_1\\ \epsilon_2\\
\epsilon_6\end{array}\right],
\end{equation}
where $P_1^0$ is the spontaneous polarization in the
$x_1$-direction. There is only one independent piezoelectric
coefficient $e_{11}$. The calculated values of the elastic and
piezoelectric constants of ZnO single layers are listed in
Table~\ref{elaconstsl}. Using the elastic constants, we calculate
the Young's modulus per atom by
$Y_{atom}=(c_{11}-c_{12}^2/c_{11})/2\approx 32.4$ eV/atom, which
is quite smaller than the value $57.7$ eV/atom for a single
graphite layer obtained in our previous work. \cite{Tuprb02}
Additionally, we observe that the ZnO single layer ($e_{11}=0.48$
e/\AA) has stronger piezoelectricity than the BN single layer
($e_{11}=0.23$ e/\AA). \cite{Saiprb03}

\section{Single-walled Zinc oxide nanotubes\label{nanotube}}
Generally speaking, if a material can exist in graphite-like
structure, the corresponding nanotubes can be synthesized in the
laboratory. \cite{Rao2003} A natural question is: how can we
fabricate SWZONTs? We propose a possible way in this section.

\begin{figure}[!htp]
\includegraphics[width=7cm]{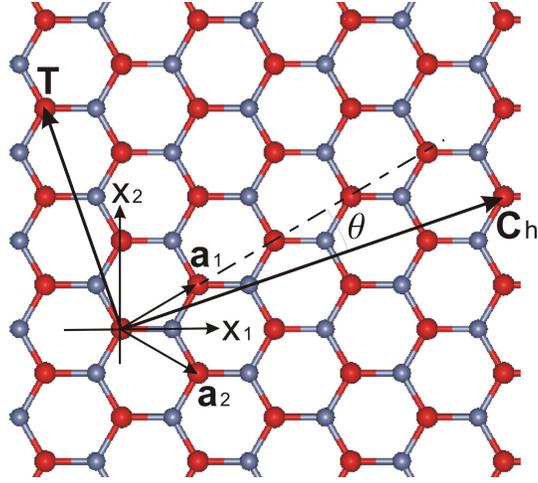}
\caption{\label{graphicly}(Color online) Unrolled honeycomb
lattice of an SWZONT. By rolling up the sheet such that the atoms
in the two ends of the vector $\textbf{C}_h$ coincide, a nanotube
is formed. The vectors $\textbf{a}_1$ and $\textbf{a}_2$ are unit
vectors of the hexagonal lattice. The translational vector
$\textbf{T}$ is perpendicular to $\textbf{C}_h$ and runs in the
direction of the tube axis.}\end{figure}

Without considering the two ends, an SWZONT can be thought of as a
cylinder rolled up from a single sheet of ZnO layer such that two
equivalent sites of the hexagonal lattice coincide. To describe
the SWZONT, we introduce several characteristic vectors in analogy
with a single-walled carbon nanotube. As shown in
Fig.~\ref{graphicly}, the chiral vector $\textbf{C}_h$, which
defines the relative location of two sites, is specified by a pair
of integers $(n, m)$ which is called the index of the SWZONT and
relates $\textbf{C}_h$ to two unit vectors $\textbf{a}_1$ and
$\textbf{a}_2$ of the hexagonal lattice ($\textbf{C}_h =
n\textbf{a}_1 + m\textbf{a}_2$). The chiral angle $\theta$ defines
the angle between $\textbf{a}_1$ and $\textbf{C}_h$. For an $(n,
m)$ nanotube, $\theta=\arccos[(2n+m)/(2\sqrt{n^2+m^2+nm})]$. The
translational vector \textbf{T} corresponds to the first point in
the hexagonal lattice through which the line normal to the chiral
vector $\textbf{C}_h$ passes.

\begin{figure}[!htp]
\includegraphics[width=7cm]{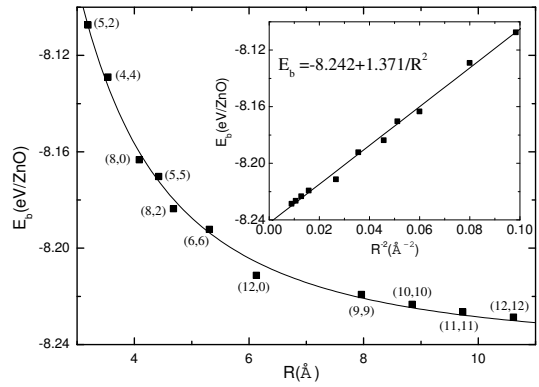}
\caption{\label{bindzont} Binding energy ($E_b$) and radius ($R$)
of different SWZONTs.}\end{figure}

We calculate the binding energy of SWZONTs with different indexes
which is shown in Fig.~\ref{bindzont}. From this figure we find
that: (i) The binding energy for different SWZONTs is negative,
which suggests these SWZONTs can exist in principle; (ii) The
binding energy ($E_b$) decreases with the increase of the radius
($R$) of SWZONTs and can be well fit by
\begin{equation}E_b=-8.242+1.371/R^2\qquad \mathrm{(eV/ZnO)}.\label{ebzont}\end{equation}
Obviously, $E_b\rightarrow -8.242$ eV/ZnO for $R\rightarrow
\infty$. This value is quite close to the binding energy (-8.246
eV/ZnO) of a ZnO single planar layer. The term $1.371/R^2$
reflects the curvature effect of nanotubes. The classic shell
theory also gives the same form, $D\Omega/R^2$, for the curvature
effect,\cite{Landauld} where $D$ is the rigidity of the ZnO single
layer and $\Omega=8.91$\AA$^2$ is the area of the parallelogram
generated by the unit vectors $\textbf{a}_1$ and $\textbf{a}_2$.
Thus we obtain $D=0.15$ eV, which is quite smaller than the
rigidity (1.17 eV)\cite{OuYang97} of a single graphite layer. The
ZnO single layer is much softer than the graphite layer such that
it should be more easily wrapped up into a nanotube.

\begin{figure}[!htp]
\includegraphics[width=7cm]{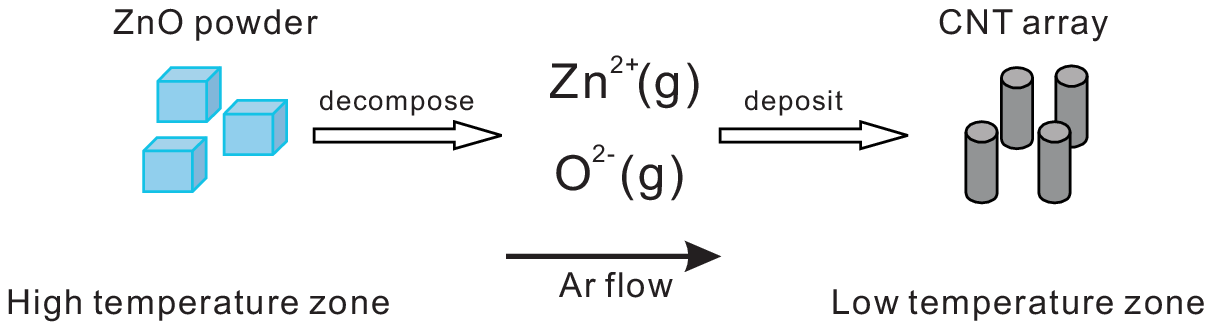}
\caption{\label{vsphase}(Color online)  Schematic diagram of the
vapor-solid phase process to synthesize SWZONTs with carbon
nanotube (CNT) array as templates.}\end{figure}

The solid-vapor phase process has been employed to successfully
synthesize ZnO nanobelts, nanorings, and nanohelixes.
\cite{Panzw,Kongxynl,Hugheswl,Kongxysci} It might also be used to
realize SWZONTs with carbon nanotubes as templates. As shown in
Fig.~\ref{vsphase}, the ZnO powder decomposes into gaseous Zn$^{2+}$
and O$^{2-}$ at high temperature and low pressure. \cite{Kohl74}
After that, Ar flow carries Zn$^{2+}$ and O$^{2-}$ to the low
temperature zone at relative higher pressure where Zn$^{2+}$ and
O$^{2-}$ will deposit on the carbon nanotube array prepared
carefully in advance. Controlling the low enough density of
Zn$^{2+}$ and O$^{2-}$ and the proper distance between carbon
nanotubes, one may obtain thin enough ZnO films, for example less
than 3 single layers, coating on the surfaces of carbon nanotubes.
These films prefer the multi-walled tubes to the wurtzite structure
because the layer number is smaller than 3. One may take out the
SWZONTs from the multi-walled tubes by analogy with the method in
carbon nanotubes. \cite{Cumingssci}

Now we discuss two points in the above process. The first point is
the growth temperature. From the conditions for producing ZnO
nanobelts, \cite{Panzw,Kongxynl,Hugheswl,Kongxysci} the growth
temperature of ZnO nanotubes is expected to be 300--500 $^\circ$C.
At this temperature, the highly purified carbon nanotubes are
chemically inactive, \cite{Ajayannat} which ensures that they merely
work as templates but do not react with Zn$^{2+}$ or O$^{2-}$. The
other point is the critical radius of the synthesized SWZONTs. The
carbon-ZnO multi-walled tubes can form in the above process if the
attraction between nanotubes overcomes the curvature energy of ZnO
nanotubes. For simplicity, we consider a carbon nanotube coated by
an SWZONT. The attraction between them is approximately described by
the Lennard-Jones potential. The layer distance is estimated to be
3--4 \AA\ from the van der waals radii of carbon, oxygen, and zinc
atoms. \cite{Partridgeh01} The attraction energy ($E_{at}$) is about
$-19.5$ meV/ZnO. \cite{remarktzc} From $E_{at}+D\Omega/R^2<0$, we
obtain the smallest radius $R_c\sim 8.3$ \AA\ synthesized in the
above process. We should emphasize that the above process is merely
possible method to produce the ZnO nanotubes in terms of the
first-principles calculation and the previous experiments
\cite{Panzw,Kongxynl,Hugheswl,Kongxysci} on the growth of ZnO
nanobelts. Whether it works or not, experimental researches will
gives an answer in the future.

It is useful to discuss the piezoelectricity of SWZONTs. We need
not to do additional DFT calculations considering the experience
obtained by Sai \emph{et al.} in the study of the piezoelectricity
of single-walled BN nanotubes. \cite{Saiprb03} They found that the
piezoelectricity of BN nanotubes depends on the chirality. The
physical origin is that an angle between local polarization of
each unit cell and the tube axis depends on the chiral angle.
Their key point is that the curvature of nanotubes has so small
effect on the piezoelectricity that we can directly deduce their
piezoelectric property from the piezoelectricity of single layers.
Two deformation modes will result in the change of the
polarization in the direction of tube axis: one is the tension or
compression strain ($\varepsilon_T$) along the tube axis, another
is the shear strain ($\gamma$, the twist angle per length) induced
by the torsion around the tube axis. The variation of the
polarization can be expressed as
\begin{equation}\delta P_T/L=e_T\varepsilon_T+e_t\gamma,\end{equation}
where $L$ represents the perimeter of the nanotube. Following
Ref.\onlinecite{Saiprb03}, one can derive the coefficients
\begin{equation}e_T=e_{11}\cos3\theta\end{equation}
and
\begin{equation}e_t=-e_{11}\sin3\theta\end{equation} from the
transformation between coordinate systems $\{x_1,x_2\}$ and
$\{\mathbf{C}_h,\mathbf{T}\}$ shown in Fig.~\ref{graphicly}. From
above three equations, we find that the piezoelectricity of
SWZONTs depends on their chiral angles. For zigzag tubes ($m=0$,
$\theta=0^\circ$), only tension or compression strains induce
polarizations, while only shear strains play roles for armchair
tubes ($n=m$, $\theta=30^\circ$). Both strains work for achiral
tubes. For a specific pure axial strain, the variation of the
polarization decreases with the increase of chiral angle $\theta$
from $0^\circ$ to $30^\circ$. For pure shear strain induced by the
torsion around the tube axis, the absolute variation of the
polarization increases with the increase of chiral angle $\theta$
from $0^\circ$ to $30^\circ$. These effects are more evident in
SWZONTs than in BN nanotubes because the ZnO single layer has a
relative larger $e_{11}$ than the BN single layer.

\section{conclusion\label{conclusion}}
We have calculated the elastic constants and piezoelectric
coefficients of ZnO crystals and single layers. We find that ZnO
single layers are much more flexible than graphite single layers in
the elasticity and stronger than boron nitride single layers in the
piezoelectricity. SWZONTs are demonstrated to exist in principle and
might be realized by the solid-vapor phase process with carbon
nanotubes as templates.

Finally, we would like to point out some potential applications of
SWZONTs and related structures. (i) SWZONTs, same as other ZnO
nanostructures, can be applied in optoelectric devices. The
quality of SWZONT-based devices should be higher than others
because of their very small thickness. (ii) The piezoelectricity
of SWZONTs depends on their chirality. For most ZnO nanotubes
except the zigzag type, twists around the tube axis will induce
axial polarizations. This property makes SWZONTs possible to be
used in the torsion measurement devices. For example, one may use
SWZONTs to make very small Coulomb torsion balance. (iii) SWZONTs,
except the armchair tubes, are polar tubes. \cite{Erkocspe} Water
transport, ice formation, and biopolymer translocation in polar
tubes should exhibit quite different characteristics from those in
non-polar tubes. The latter has been fully studied, especially for
carbon nanotubes, \cite{Hummernat,koganature,Gaohnl03} while the
former has attracted no attention even for BN tubes up to now.
(iv) The carbon-ZnO nanotubes formed in the solid-vapor phase
process must have excellent mechanical and electrical properties:
On the one hand, they have high enough rigidity and axial strength
because of the contribution of carbon nanotubes. On the other
hand, they exhibit strong piezoelectricity coming from the ZnO
tubes. These structures may be employed to make electric nanotube
motors \cite{Tuprb05} where the axial voltage will be greatly
reduced relative to carbon nanotube motors \cite{Tuprb05} because
of the high piezoelectricity of ZnO tubes.

\section*{Acknowledgements}
We are very grateful to the kind help of Dr. Q. X. Li and Y. D.
Liu. Calculations have been performed on Numerical Materials
Simulator (HITACHI SR11000) at the Computational Materials Science
Center, National Institute for Materials Science.

\begin{table}[!htp]
\caption{Structure parameters of ZnO crystals.\label{strutpara}}
\begin{ruledtabular}
\begin{tabular}{cccccc}
Authors & Methods & $a$ (\AA) & $c$ (\AA)& $c/a$ & u\\ \hline
Present & DFT & 3.199 & 5.167 & 1.615 & 0.379\\
Karzel \emph{et al.} \footnote{Reference \onlinecite{Karzelprb}.}&
Experiment & 3.250 & 5.204 &1.602 & 0.382\\
Wu \emph{et al.} \footnote{Reference \onlinecite{wuxfprb}.}& DFT &3.197 & 5.166 & 1.616 & 0.380\\
Catti \emph{et al.} \footnote{Reference \onlinecite{Cattijpcs}.}&
Hartree-Fock & 3.286 & 5.241 & 1.595 & 0.383
\end{tabular}
\end{ruledtabular}
\end{table}

\begin{table}[!htp]
\caption{Elastic constants (relaxed-ions) of ZnO crystals in units
of GPa.\label{elaconst}}
\begin{ruledtabular}
\begin{tabular}{ccccccc}
Authors & Methods & $c_{11}$ & $c_{12}$ & $c_{13}$ & $c_{33}$ & $c_{44}$\\
\hline
Present & DFPT & 218 & 137 & 121 & 229 & 38\\
Wu \emph{et al.} \footnote{Reference \onlinecite{wuxfprb}.}& DFPT & 226 & 139 & 123 & 242 & 40\\
Catti \emph{et al.} \footnote{Reference \onlinecite{Cattijpcs}.}&
Hartree-Fock & 246 & 127 & 105 & 246 & 56\\
Kobiakov\footnote{Reference \onlinecite{Kobiakovssc}.}&Experiment
&207&118&106&210&45\\
Azuhata \emph{et al.}\footnote{Reference
\onlinecite{Azuhatajap}.}&Experiment &190&110&90&196&39
\end{tabular}
\end{ruledtabular}
\end{table}

\begin{table}[!htp]
\caption{Piezoelectric coefficients (relaxed-ions) of ZnO crystals
in units of C/m$^2$.\label{piezoconst}}
\begin{ruledtabular}
\begin{tabular}{ccccc}
Authors & Methods & $e_{31}$ & $e_{33}$ & $e_{15}$ \\
\hline
Present & DFPT & -0.65 & 1.24 & -0.54 \\
Wu \emph{et al.} \footnote{Reference \onlinecite{wuxfprb}.}& DFPT & -0.67 & 1.28 & -0.53\\
Catti \emph{et al.} \footnote{Reference \onlinecite{Cattijpcs}.}&
Hartree-Fock & -0.55 & 1.19 & -0.46 \\
Kobiakov\footnote{Reference \onlinecite{Kobiakovssc}.}&Experiment
&-0.62&0.96&-0.37
\end{tabular}
\end{ruledtabular}
\end{table}

\begin{table}[!htp]
\caption{Optimized structure of a ZnO single layer. Here $L_v$,
$B_l$, $B_a$, and $E_b$ represent the vacuum layer thickness, bond
length, bond angle, and binding energy, respectively.
\label{planartab}}
\begin{ruledtabular}
\begin{tabular}{ccccc}
$L_v$ (\AA) & ecut (Ha) & $B_l$ (\AA) & $B_a$ & $E_b$ (eV/ZnO)\\
\hline
10.6 & 35 &  1.852 & 120$^{\circ}$ & -8.246\\
21.2 & 60 &  1.852 & 120$^{\circ}$ & -8.247
\end{tabular}
\end{ruledtabular}
\end{table}

\begin{table}[!htp]
\caption{Elastic and piezoelectric  constants (relaxed-ions) of
ZnO single layers. Here $L_v$ represents the vacuum layer
thickness.\label{elaconstsl}}
\begin{ruledtabular}
\begin{tabular}{ccccc}
$L_v$ (\AA) & ecut (Ha) & $c_{11}$ (eV/ZnO) & $c_{12}$ (eV/ZnO) & $e_{11}$ (e/\AA)\\
\hline
10.6 & 35 &  72.6 & 24.0 & 0.48 \\
21.2 & 60 &  72.7 & 24.1 & 0.48
\end{tabular}
\end{ruledtabular}
\end{table}

\newpage

\begin{thebibliography}{99}
\bibitem{remarktzc}The characteristic energy of van der Waals
interactions (in Lennard-Jones form) between cabon atoms, oxygen
atoms, and zinc atoms is taken from
Ref.~\onlinecite{Partridgeh01}: $\varepsilon_{C}=-24.95$ K,
$\varepsilon_{O}=-22.75$ K, $\varepsilon_{Zn}=-111.37$ K. Using
the combination law of Lennard-Jones potentials, we obtain the
characteristic energy of interactions between carbon and oxygen
atoms, and carbon and zinc atoms as
$\varepsilon_{C-O}=-(\varepsilon_{C}\varepsilon_{O})^{1/2}=-2.0$
meV,
$\varepsilon_{C-Zn}=-(\varepsilon_{C}\varepsilon_{Zn})^{1/2}=-4.5$
meV. Roughly speaking, each zinc and oxygen atoms interact with
three nearest neighbor carbon atoms. The attraction energy
($E_{at}$ per ZnO) between ZnO tube and carbon tube is estimated
as $E_{at}=-3(\varepsilon_{C-Zn}+\varepsilon_{C-O})$.
\end{thebibliography}
\end{document}